\documentclass{phostproc}

\usepackage{lipsum} 

\usepackage[table, dvipsnames]{xcolor}

\title{Model physics in low-mass solar-type stars: atomic diffusion and metallicity mixture}
\author{Benard Nsamba,$^{1,2}$
        Tiago L. Campante,$^{1,2}$
        M\'ario J. P. F. G. Monteiro,$^{1,2}$
        Margarida S. Cunha, $^{1,2}$ \\  
        Ben M. Rendle, $^{3,4}$
        Daniel R. Reese, $^{5}$
        and Kuldeep Verma $^{4}$
 }

\affiliation{
$^{1}$Instituto de Astrof\'{\i}sica e Ci\^{e}ncias do Espa\c{c}o, Universidade do Porto,  Rua das Estrelas, PT4150-762 Porto, Portugal\\
$^{2}$Departamento de F\'{\i}sica e Astronomia, Faculdade de Ci\^{e}ncias da Universidade do Porto, Rua do Campo Alegre, s/n, PT4169-007\\ Porto, Portugal\\
$^{3}$School of Physics and Astronomy, University of Birmingham, Edgbaston, Birmingham B15 2TT, UK\\
$^{4}$Stellar Astrophysics Centre, Department of Physics and Astronomy, Aarhus University, Ny Munkegade 120, DK-8000 Aarhus C, Denmark\\
$^{5}$LESIA, Observatoire de Paris, Universit\'{e} PSL, CNRS, Sorbonne Universit\'{e}, Univ. Paris Diderot, Sorbonne Paris Cit\'{e}, 5 place Jules Janssen, 92195 Meudon, France
}

\shorttitle{Systematics from input physics}
\shortauthors{B. Nsamba \textit{et al.}} 

\abs{
Using asteroseismic data from the {\it Kepler} satellite, we explore the systematic uncertainties arising from changes in the input physics used 
when constructing evolution models of solar-type stars. We assess the impact of including atomic diffusion and of varying the metallicity mixture on the determination of global stellar parameters (i.e., radius, mass, and age). We find significant systematic uncertainties on global stellar parameters when diffusion is included in stellar grids. Furthermore, we find the systematic uncertainties on the global stellar parameters to be comparable to the statistical uncertainties when a different metallicity mixture is employed in stellar grids.
}

\begin{document}

\maketitle

\section{Introduction}

Stellar model physics is known to play an important role in the evolution process and position of stars across the Hertzsprung–Russell diagram.
It also plays a vital role towards the characterisation of stars and understanding their interiors.
In preparation for recently launched NASA’s Transiting Exoplanet Survey Satellite (TESS; \citealt{Campante}) and forthcoming ESA’s
PLAnetary Transits and Oscillations of stars (PLATO; \citealt{Rauer1}) mission, we find it relevant to explore the systematic uncertainties 
on global stellar parameters (i.e., radius, mass, and age) that arise from the physics used in stellar models.

Atomic diffusion is known to be an important process in low-mass stars (e.g., \citealt{Valle2014, Valle2015,Dotter2017}), and we explore the systematic
uncertainties arising from its inclusion in stellar grids. Furthermore, we highlight the impact of 
the uncertainty in the metallicity mixture and quantify the systematic uncertainties induced on the global stellar parameters.

This  article is  organised  as  follows.  In  Sect.~\ref{target},   we describe our target stars,  seismic and classical constraints, and provide a description
of our model grids. In Sect.~\ref{results}, we summarise our results and conclusions.

\section{Target stars and model grids}
\label{target}

\begin{figure}[!h]
	\centering
	\includegraphics[width=0.85\linewidth]{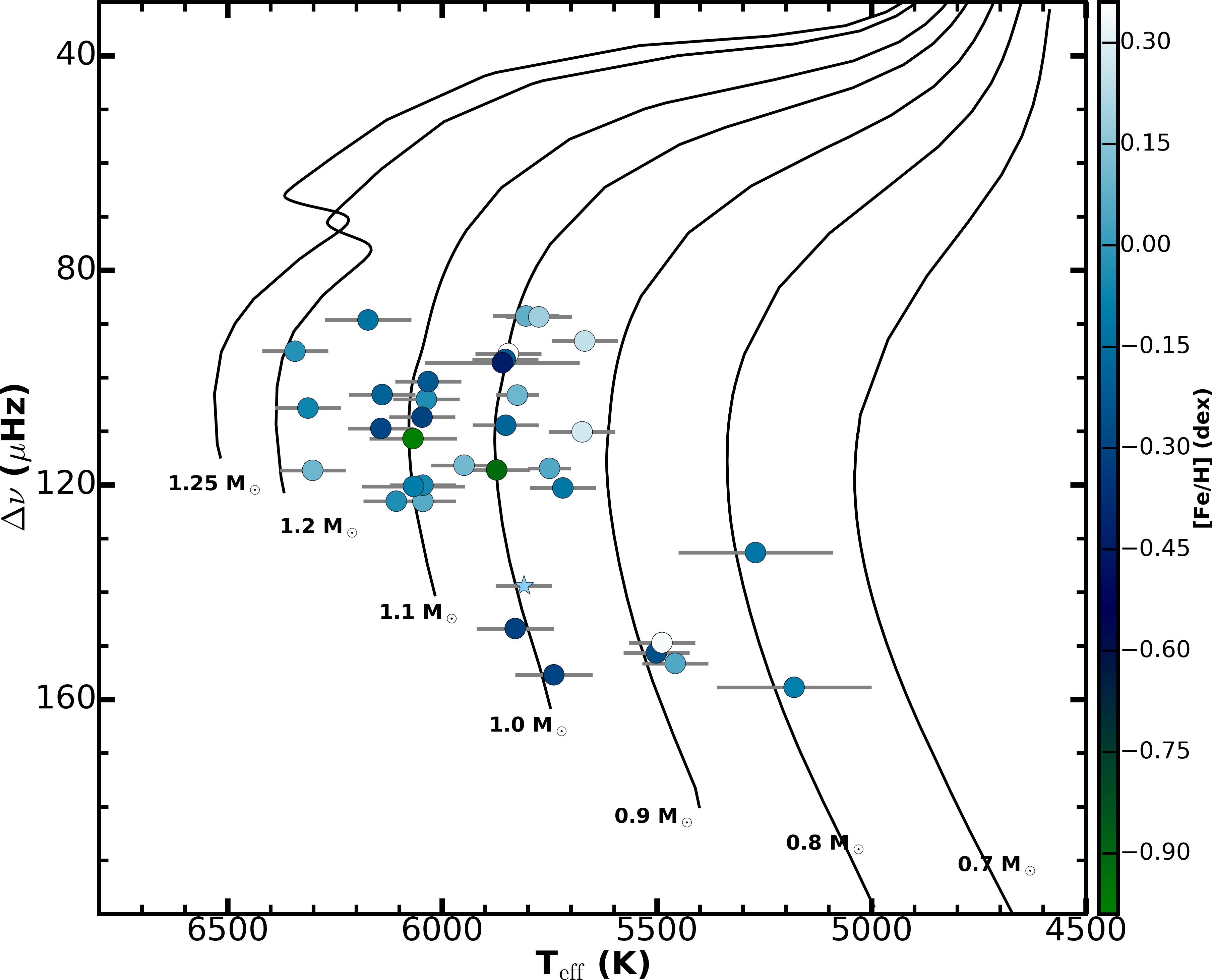}
	\caption{Evolution tracks constructed at solar metallicity and ranging in mass from 0.7 to 1.25 M$_\odot$. Stars are colour-coded
according to their metallicity. The ``star'' symbol corresponds to the position of the Sun.}
	\label{sample}
\end{figure}

Our sample consists of the 34 low-mass (i.e., below 1.2 M$_\odot$), solar-type stars with {\it Kepler} photometry (\citealt{Borucki}) shown in Fig.~\ref{sample}. These stars have high S/N in the oscillations. Individual oscillation frequencies 
for each star in the sample are adopted from \citet{Lund}, while spectroscopic constraints (i.e., effective temperature, $T_{\rm eff}$, and metallicity, [Fe/H]) are from \citet{Aguirre1} and references therein.

Using the stellar evolution code {\sc mesa} (Modules for Experiments in Stellar Astrophysics; \citealt{Pax3}), we set up three grids varying only in the specific model physics 
being investigated (see Table \ref{constituents}). The evolution tracks vary in mass, $M \in$ [0.70, 1.25]M$_\odot$ in steps of 0.05, initial metal mass fraction, $Z_0$  $\in$ [0.006, 0.031]
in steps of 0.001, and mixing length parameter, $\alpha_{\rm mlt}$ $\in$ [1.3, 2.9] in steps of 0.1.
For further details on the grid properties, such as the nuclear reaction rates tables, equation of state, opacities, model atmosphere tables etc., please 
see \citet{Nsamba2018}.

We note that the initial helium mass fraction of our evolution models was determined using the helium-to-heavy-metal enrichment law, expressed as
\begin{equation}
 Y = \left(\frac{\Delta Y}{\Delta Z}\right)Z + Y_0 \, ,
	\label{mass_fraction}
\end{equation}
with $Y_0$ set to the big bang nucleosynthesis primordial value of 0.2484 when $Z$ = 0.0 (\citealt{Cyburt}) and $\Delta Z / \Delta Y$ = 2. The treatment of $Y$ is expected to be 
a significant source of systematic uncertainty on the stellar properties and is currently being addressed in Nsamba et al.~(in prep.).

\begin{table*}[!h]
	\centering
	\caption{Summary of the main model grid properties.}
	\label{tab:table_wide}
	\begin{tabular*}{\linewidth}{l @{\extracolsep{\fill}} c l c ll}
	\noalign{\smallskip}\hline\hline\noalign{\smallskip}
	     Name & Mass (M$_\odot$) & Solar mixture & $\frac{\Delta Y}{\Delta Z}$ & Overshooting & Diffusion\\
	\noalign{\smallskip}\hline\noalign{\smallskip}
	GS98{\small sta} &  0.70 - 1.25 &  \citet{Grevesse} &  2.0  & No & Yes \\
	GS98{\small nod} &  0.70 - 1.25 &  \citet{Grevesse} &  2.0  & No & No \\
	AGS09		 &  0.70 - 1.25 &  \citet{Asplund}   &  2.0  & No & Yes \\
	\noalign{\smallskip}\hline
	\end{tabular*}
	\label{constituents}
\end{table*}

Adiabatic pulsation frequencies for each evolution model were calculated using {\sc gyre} (\citealt{Townsend}) for spherical degrees $l$ = 0, 1, 2, and 3. The surface effect is corrected using the two-term surface correction by \citet{Ball} implemented in {\sc aims} (Asteroseismic Inference on a Massive Scale; \citealt{Reese}, Rendle et al.~submitted). Stellar parameters and their corresponding uncertainties  are obtained from the statistical mean and standard deviation of the posterior probability density functions (PDFs) returned by {\sc aims}.

\section{Results and conclusions}
\label{results}

\begin{figure}[!h]
	\centering
	\includegraphics[width=0.85\linewidth]{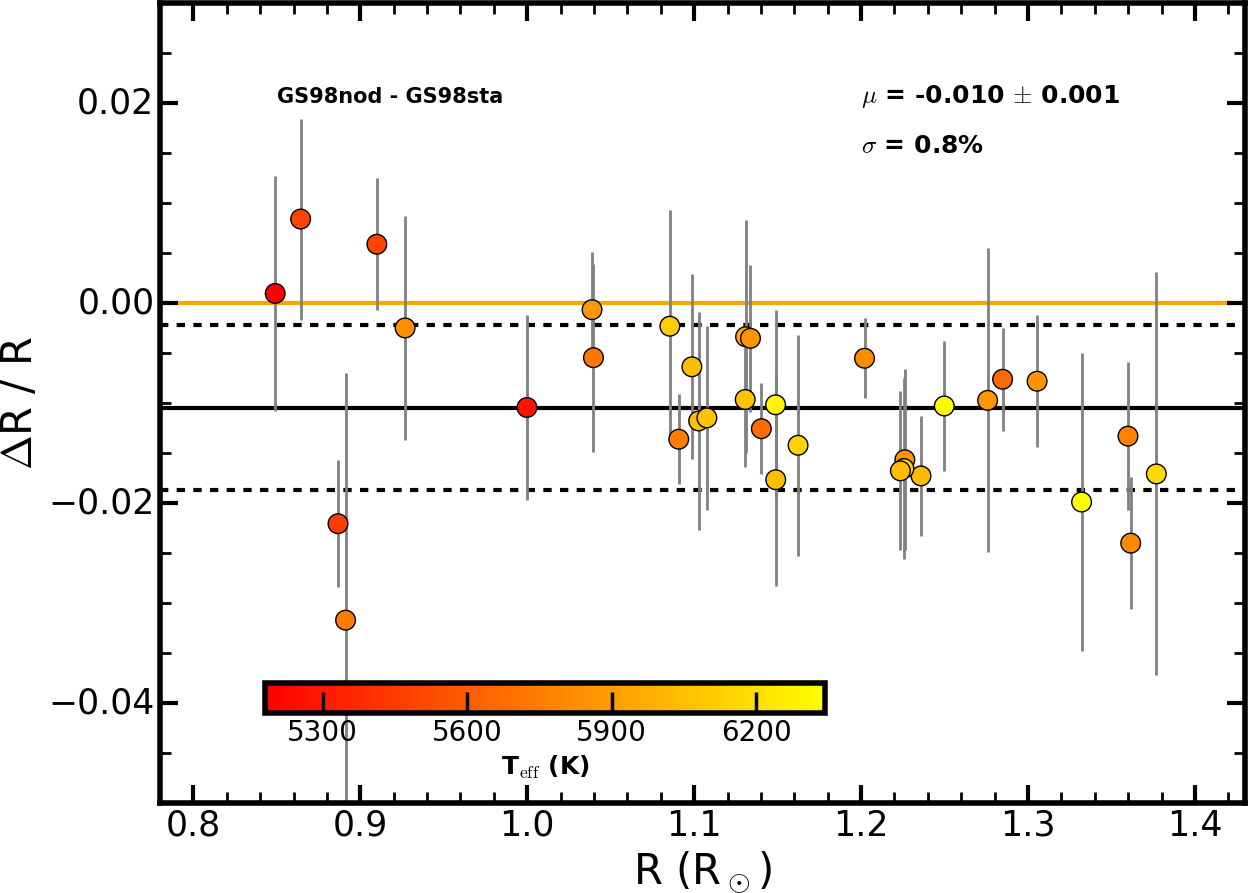}\\
	\vspace{0.4cm}
	\includegraphics[width=0.85\linewidth]{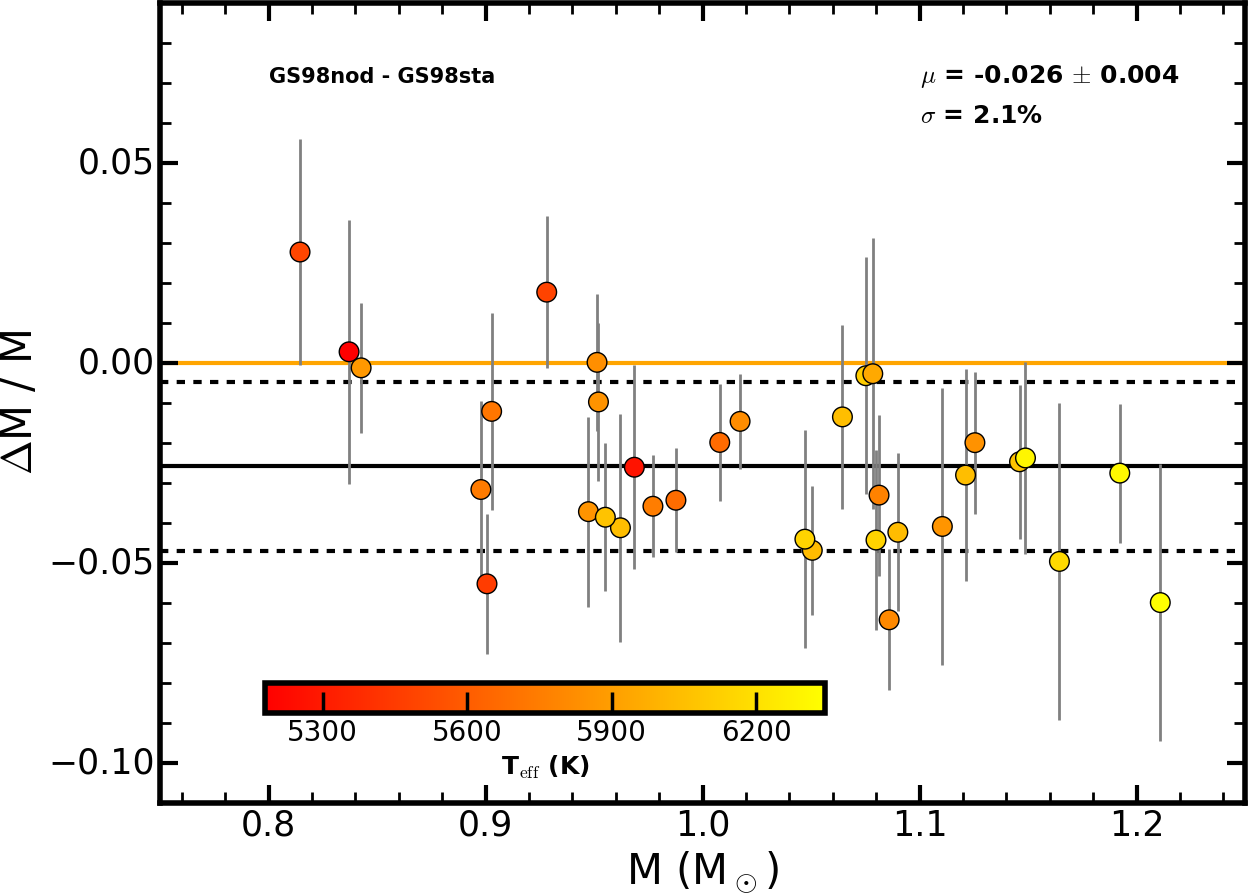}\\
	\vspace{0.4cm}
	\includegraphics[width=0.85\linewidth]{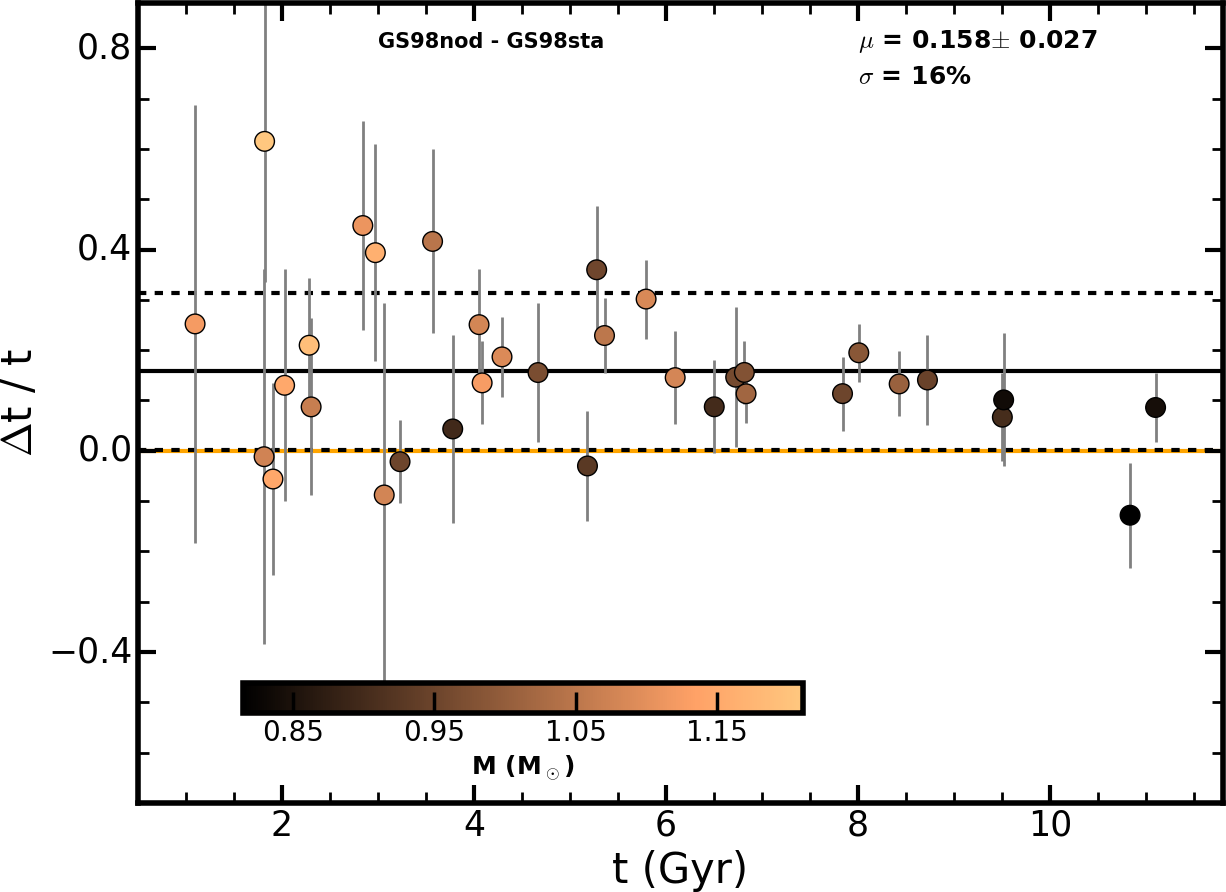}\\
	\caption{Fractional differences in stellar radius, mass, and age resulting from the inclusion of diffusion (abscissa values are from GS98sta). GS98nod corresponds to the grid without diffusion. The orange line is the null offset, the black solid line represents the bias ($\mu$), and the scatter ($\sigma$) is represented by the dashed lines.}
	\label{diff}
\end{figure}

The systematic uncertainties arising from the inclusion of atomic diffusion are shown in Fig.~\ref{diff}. They amount to 0.8\%, 2.1\%, and 16\% in radius, mass, and age, respectively. The lower panel of Fig.~\ref{diff} shows that stellar ages computed based on the grid with diffusion are on average lower than those from the grid without diffusion. The systematic uncertainties in stellar mass and age are significantly larger than the corresponding
statistical uncertainties. For a comprehensive discussion, please refer to \citet{Nsamba2018}.
It is interesting to note that the mass and age seem to be anti-correlated (see Fig.~\ref{diff}) as expected from stellar evolution.

\begin{figure}[!h]
	\centering

	\includegraphics[width=0.85\linewidth]{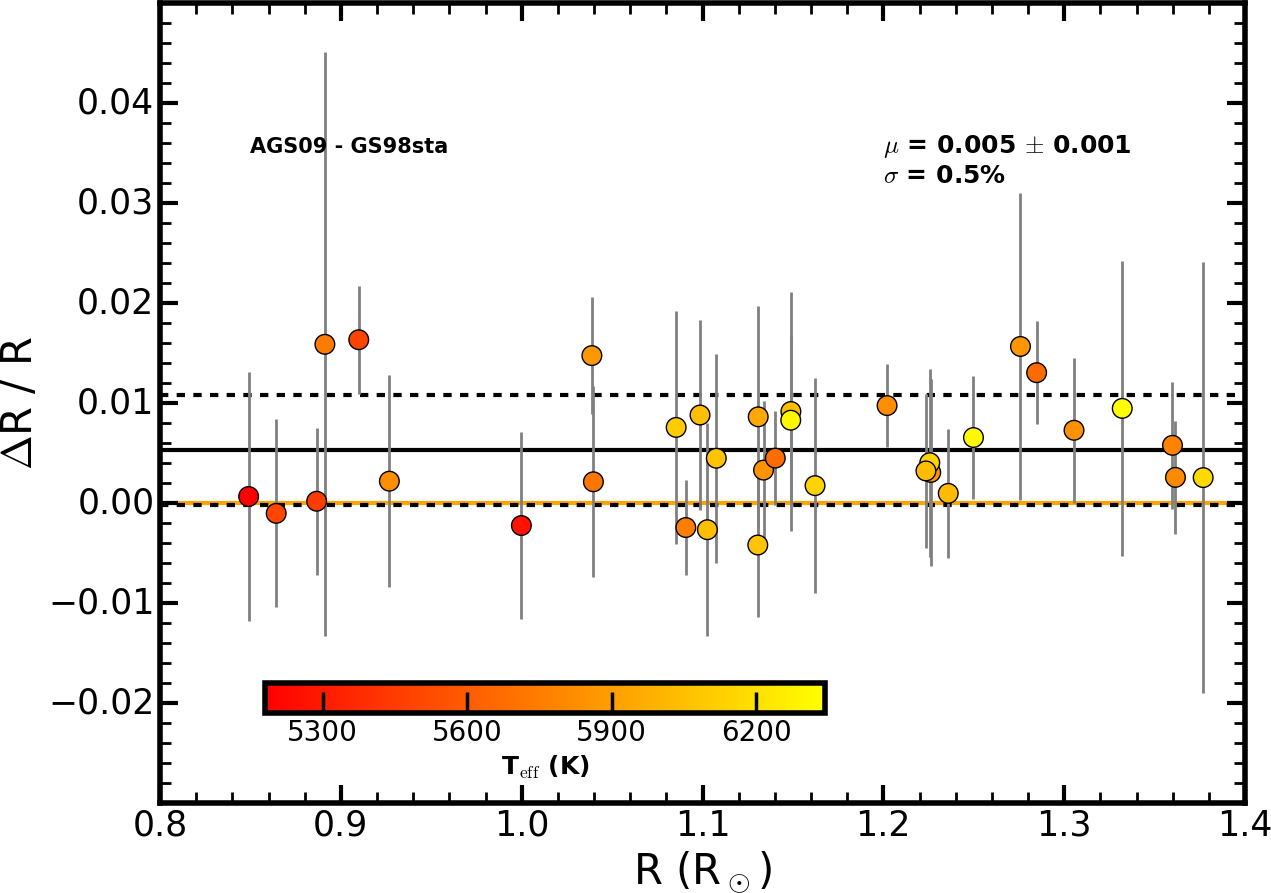}\\
	\vspace{0.4cm}
	\includegraphics[width=0.85\linewidth]{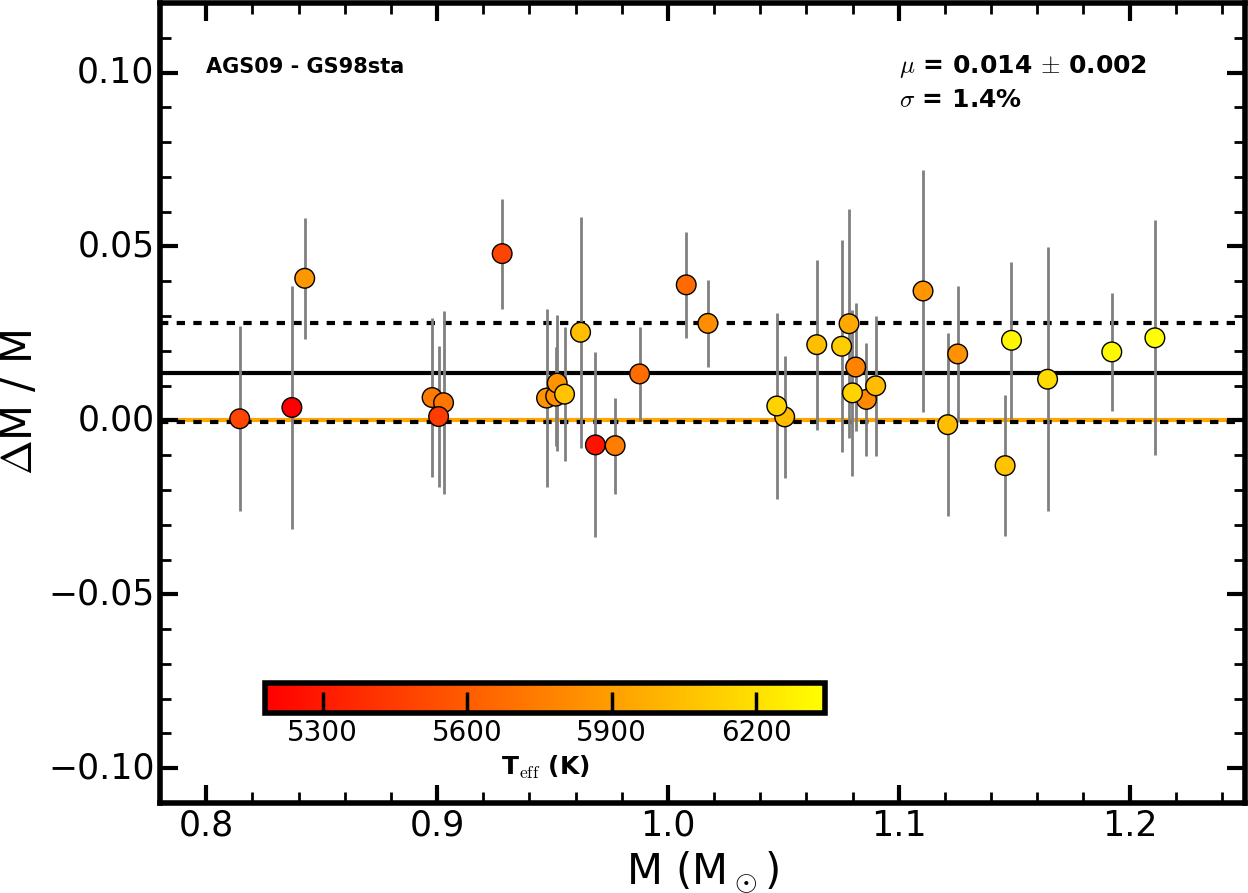}\\
	\vspace{0.4cm}
	\includegraphics[width=0.85\linewidth]{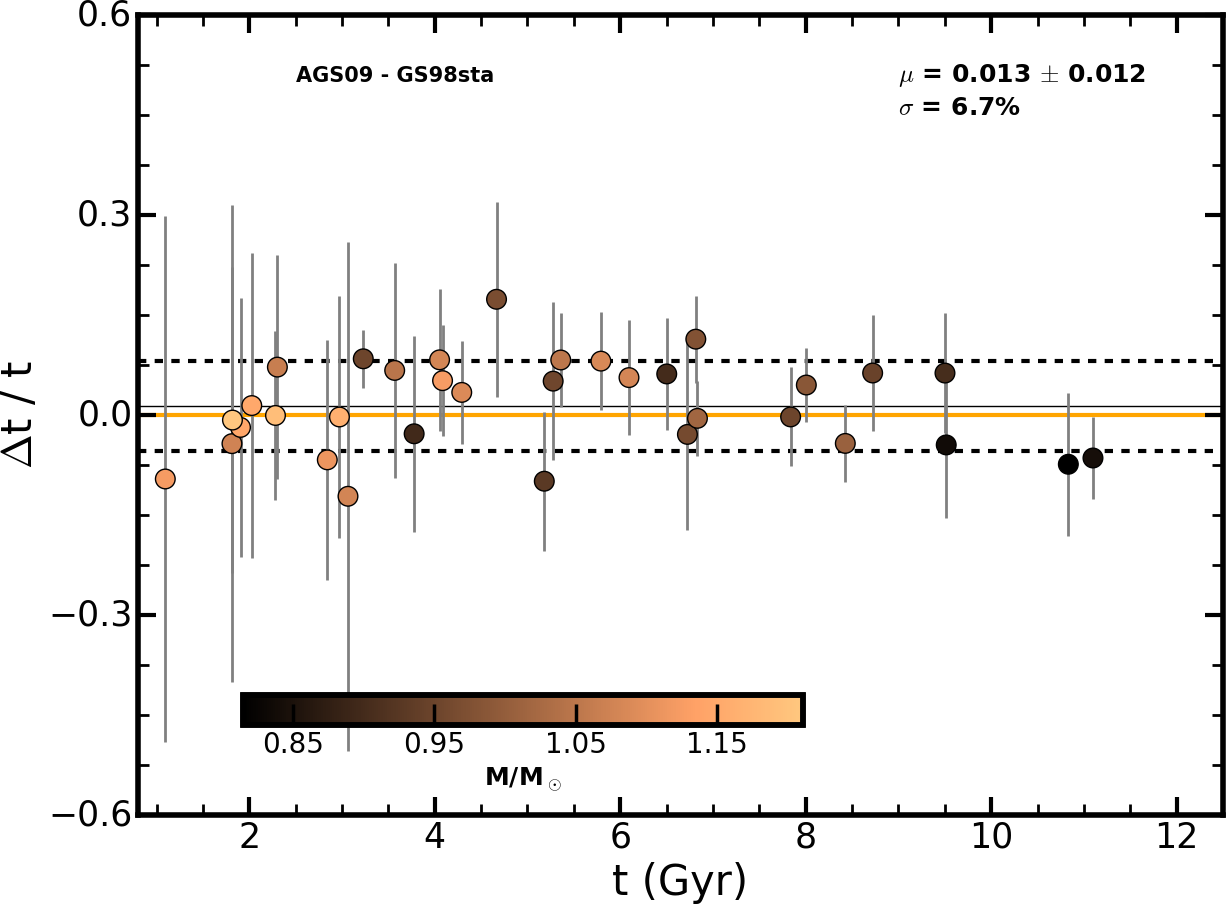}\\
	\caption{Fractional differences in stellar radius, mass, and age resulting from the adoption of a different metallicity mixture (abscissa values are from GS98sta). The orange line is the null offset, the black solid line represents the bias ($\mu$), and the scatter ($\sigma$) is represented by the dashed lines.}
	\label{comp}
\end{figure}

The systematic uncertainties arising from the adoption of a different metallicity mixture are shown in Fig.~\ref{comp}. They amount to 0.5\%, 1.4\%, and 6.7\% in radius, mass, and age, respectively. The statistical uncertainties are comparable to the systematic uncertainties in this case, consistent with the findings of \citet{Aguirre}.

In conclusion, we find that atomic diffusion plays a vital role in the input physics with regard to low-mass, solar-type stars, its impact being significant on the computed mass and age. Our findings also show that the uncertainty in the metallicity mixture has a limited impact on the global stellar parameters. Note that variation of the metallicity mixture implies setting the appropriate opacities during grid construction. Therefore, the systematic uncertainties found are from both these inputs. We refer to \citet{Nsamba2018} for further details on the discussion of the model physics highlighted here, including the systematic uncertainties on the global stellar parameters arising from different surface correction methods.

\section*{Acknowledgements}
This work was supported by FCT - Fundação para a Ciência e a Tecnologia  through national funds and by FEDER through COMPETE2020 - Programa Operacional Competitividade e Internacionalização by these grants: UID/FIS/04434/2013 \& POCI-01-0145-FEDER-007672, PTDC/FIS-AST/30389/2017 \& POCI-01-0145-FEDER-030389 and PTDC/FIS-AST/28953/2017 \& POCI-01-0145-FEDER-028953.
 B.~Nsamba is supported by Funda\c{c}\~{a}o para a Ci\^{e}ncia e a Tecnologia (FCT, Portugal) under the Grant ID: PD/BD/113744/2015 from PhD::SPACE, an FCT PhD program. T.~L.~Campante acknowledges support from the European Union’s Horizon 2020 research and innovation program under the Marie Sk\l{}odowska-Curie grant agreement No.~792848 and from grant CIAAUP-12/2018-BPD.

\bibliographystyle{phostproc}
\bibliography{input_physics.bib}

\end{document}